\documentclass[a4paper,11pt]{article}
\pdfoutput=1 

\usepackage{jcappub} 
\usepackage{booktabs}
\usepackage[english]{babel}
\usepackage{amsmath,amssymb,amsbsy,amstext, amsthm, simplewick}
\usepackage{hyperref}
\usepackage{graphicx}
\usepackage{amsfonts}
\usepackage{amssymb}
\usepackage[small]{caption}
\usepackage{upgreek}
\usepackage{cancel}
\usepackage{color}
\usepackage{slashed}
\usepackage[dvipsnames]{xcolor}
\usepackage{subcaption}
\usepackage{tikz-cd}
\usepackage{feynmp-auto}
\usepackage{color}
\usepackage{soul}
\usepackage{dsfont}
\usepackage{verbatim}

\newcommand{\half}{\frac{1}{2}}

\usepackage{empheq}
\newlength\dlf  

\title{Neutrino -- Dark Matter Scattering and Coincident Detections of UHE Neutrinos with EM Sources}
\author{Seth Koren}
\affiliation{Department of Physics, University of California, \\
Santa Barbara, CA 93106, U.S.A.}

\emailAdd{koren@physics.ucsb.edu}

\abstract{
The scattering of neutrinos off dark matter can induce time delays in their propagation compared to that of photons, which would wash out correlations between ultra-high-energy neutrinos and electromagnetic observations of their sources --- while preserving the observed diffuse neutrino flux. This may explain the significant discrepancy between predictions of neutrino fluxes from gamma ray bursts and the lack of neutrinos correlated with EM observations of GRBs. Conversely, the detection of an UHE neutrino in association with a source provides a strong constraint on such interactions. We consider an effective model of dark photon dark matter interacting with neutrinos which exhibits this effect. 
}
\arxivnumber{1903.05096}

\begin{document}
\maketitle

\section{Introduction} \label{sec:intro}

Searches for dark matter have been conducted for decades by astronomers and particle physicists alike. In a field so mature with as yet so many unknowns, any new observational window or hint of new physics deserves consideration. In this work we consider an unlikely source of new information about dark matter: Associations of observed ultra-high-energy neutrinos with astrophysical sources. 

With the continuing null results of searches for traditional WIMPs, attention has increasingly been diverted toward dark matter candidates which are neutral under the SM gauge group. The SM offers three renormalizable ways that entirely dark sectors may nonetheless communicate with the SM fields, one of which is the `neutrino portal': a singlet fermion may have a Yukawa coupling with the Higgs and the SM leptons. Interactions through this portal are difficult to probe directly, which makes the neutrino sector an intriguing cranny wherein dark matter interactions with the SM may be hiding. But neutrino phyiscs, while secluded, is not entirely hidden. As measurements of terrestrial and solar neutrinos have progressed dramatically, so too has our ability to detect astrophysical neutrinos. And along with the impressive measurements of neutrinos of extragalactic origin a new enigma has arisen.

The existence of ultra-high energy (UHE) cosmic rays has been a persistent source of mystery in astrophysics, and much effort has been expended on understanding possible sources both theoretically and observationally. The cosmic ray components most easily-studied are those which are electrically charged --- but as a result these particles have only limited ability to tell us about their sources, as they are buffeted about by intergalactic magnetic fields on their journey to Earth.

However, it has long been theorized that there is a neutrino component to cosmic rays. In the hot, dense environments which accelerate protons to ultra-high energies, neutrinos may be produced through the creation of pions by $p\gamma$ interactions. The cosmic presence of UHE neutrinos was confirmed by IceCube in \cite{Aartsen:2013jdh,Aartsen:2014gkd} with the detection of an (approximately) isotropic, flavor-universal\footnote{Flavor-universality is consistent with production by pion decay after accounting for the results of mixing over cosmological distances. See \cite{Aartsen:2015ivb} and references therein.} flux of UHE neutrinos with energies $30 \text{ TeV} - 2 \text{ PeV}$. In the Standard Model, neutrinos interact only weakly and so are unlikely to interact with the intergalactic medium on their way to us. They thus offer the prospect of determining the sources of cosmic rays, and this idea has long spurred neutrino astronomy. 

Gamma-ray bursts (GRBs) have been a leading theoretical explanation for the source of these neutrinos \cite{Waxman:1997ti,Vietri:1998nm,Halzen:2002pg,Meszaros:2015krr}. 
Yet despite sensitive searches for UHE neutrinos in association with an observed GRB, searches have come up empty \cite{Aartsen:2016qcr,Aartsen:2017wea,Albert:2016eyr,2012Natur.484..351A,2010ApJ...710..346A,2015ApJ...805L...5A,2007ApJ...664..397A,2008ApJ...674..357A,Adrian-Martinez:2013dsk}. Large astrophysical uncertainties are involved in the theorized neutrino spectra, and so predictions for the peak neutrino energy vary between 100 TeV and 1 EeV. The initial hope of up to ${\sim} 100/\text{yr}$ UHE neutrinos coincident with GRBs has been revised downward by more recent models, but the observational result of $\lesssim 1 / 5 \text{yr}$ is now in significant tension with the hypothesis of UHE neutrino production in GRBs \cite{He_2012,Zhang:2012qy,Gao:2013fra,Baerwald:2014zga,Yacobi:2014vja,Guetta:2015iza,Hooper:2018wyk,Aartsen:2016qcr,Aartsen:2017wea,2012Natur.484..351A}.

The trailblazing observation IceCube-170922A of an UHE neutrino by IceCube in ${\sim} 3 \sigma$ coincidence with a blazar outburst \cite{IceCube:2018dnn} heralds a new era of neutrino astrophysics (see \cite{Ackermann:2019cxh,Ackermann:2019ows} for recent review of near-future prospects). But tensions in accommodating this association in simple blazar models have been noted \cite{Keivani:2018rnh,Hooper:2018wyk,Gao:2018mnu}, especially due to the fact that prior neutrino detections from its direction occurred during quiescent periods \cite{IceCube:2018cha,Wang:2018zln,Rodrigues:2018tku}. And if the association is physical, a variety of complementary observations constrain such bright blazar sources to be responsible for only a small subcomponent of the UHE neutrino flux \cite{Aartsen:2016lir,Murase:2018iyl,Palladino:2018lov,Hooper:2018wyk}, which makes it a surprising contender for the first coincident detection. In the coming years similar observations --- or the lack thereof --- will provide the final word on whether we can detect UHE neutrinos in association with their sources. In the meantime, we consider both possibilities and understand the possible interplay between these observations and neutrino interactions with dark matter.

On the one hand, if this association is purely coincidental then we have not seen a single UHE neutrino associated with a source. We will show that this may be a result of UHE neutrinos being scattered away from our line of sight via interactions with dark matter, which would explain our inability to see correlated sources while still allowing for the presence of the diffuse background. On the other hand, if the association of an IceCube neutrino with the TXS 0506+056 blazar event is physical, this single event places tight constraints on models in which light dark matter interacts with neutrinos. This has also been observed recently in \cite{Kelly:2018tyg,Huang:2018cwo,Choi:2019ixb,Alvey:2019jzx,Murase:2019xqi}. We here consider an effective model of neutrinos interacting with dark photon dark matter, and show that this model may either explain the missing GRB neutrinos or be constrained by the blazar event. We describe briefly a smoking gun signature of neutrino -- dark matter scattering which may be used to confirm its presence despite uncertainties on the source spectra.

This paper is organized as follows. In Section \ref{sec:requirements} we calculate the scattering rate required to decorrelate observations of UHE neutrinos and their sources. This guides us in Section \ref{sec:model} to consider an effective model of light dark photon dark matter interacting with neutrinos. In Section \ref{sec:constraints} we consider complementary effects of such interactions and constraints thereon. We conclude in Section \ref{sec:conclusion}.

\section{General Considerations} \label{sec:requirements}

Since astrophysical uncertainties on the fluxes of neutrinos produced by high-energy sources are large, our bounds will necessarily be order-of-magnitude. Our physical criterion is that if most neutrinos are scattered on their way to us we should never see any coincident with sources. We will use a benchmark inspired by the characteristics of IceCube-170922A/TXS 0506+056, of neutrino energy $E_\nu \sim 300 \text{ TeV}$ and source distance $d \sim 1 \text{ Gpc}$. 

We here estimate the cross-section required for significant scattering under the assumption that any scattering event removes an emitted neutrino from our line of sight. Neutrinos will be scattered into our line of sight from some other solid angle --- but over cosmological distances these will reach us with a significant time delay and no longer be observationally associated with their source. The window in which IceCube looks for coincident neutrinos, which is taken from the duration of electromagnetic events, is typically $\Delta t \sim 10^3 \text{ s}$ at most \cite{Aartsen:2016qcr,Aartsen:2017wea}. The time delay of an incident neutrino due to a single scattering event is $c \Delta t \sim \theta^2 d$, so a scattering angle of $\theta \gtrsim 10^{-7}$ (in the lab frame) is required for such a scatter to decorrelate neutrino and electromagnetic observations. 

For a source a distance $d$ away, the fraction of neutrinos that reach us is $e^{-d/\lambda}$, where $\lambda(E_\nu)$ is the mean free path of an UHE neutrino in the dark matter background. In the approximation that dark matter is homogeneously distributed, the mean free path is a constant $\lambda = 1/\sigma n$ with $n$ the average number density of dark matter and $\sigma(E_\nu)$ the total scattering cross-section. On cosmological scales the average energy density of cold dark matter is measured to be $\Omega_c \rho_{crit} = \rho_{CDM} \sim 10^{-30} \text{g/cm}^3 \sim 10^{-47} \text{ GeV}^4$ \cite{Aghanim:2018eyx}, and so its number density is $n \simeq \rho_{CDM} /m_X$, where $m_X$ is the mass of the dark matter.\footnote{Observations are consistent with an $\mathcal{O}(1)$ fraction of dark matter being unbound to galaxies, and so comprising a dark intergalactic medium. See \cite{Karachentsev:2018ysz} for a recent discussion and review. We here use the simplifying assumption of complete homogeneity in calculating a benchmark cross-section for neutrino -- dark matter scattering, but the unknown dark matter distribution in intergalactic space and the unknown positions of UHE neutrino sources within galactic dark matter halos are sources of uncertainty.} An approximate dividing line for whether attenuation has an appreciable effect is $\lambda \gtrless d$, which translates into

\begin{equation}
 \sigma(E_\nu) \lessgtr 10^6  \text{ GeV}^{-2} \left(\frac{m_X}{\text{GeV}}\right) \left(\frac{\text{Gpc}}{d}\right).
\end{equation}

Upon association of an UHE neutrino of energy $E_\nu$ with a source a distance $d$ away, this is an approximate upper bound on the scattering cross-section of neutrinos with dark matter of mass $m_X$. Conversely, this is an approximate lower limit on the cross-secton required for neutrino -- dark matter scattering to explain the dearth of UHE neutrinos of typical energy $E_\nu$ correlated with sources at typical distances $d$. For our benchmark event we define a benchmark cross-section at $E_\nu = 300 \text{ TeV}$ of $\sigma_0(m_X) \equiv 10^6 \text{ GeV}^{-3} m_X$. 

The unitarity limit on $2 \rightarrow 2$ scattering, assuming it's dominated by low $\ell$ partial waves, is roughly $\sigma \lesssim \frac{8 \pi}{m_X E_\nu}$ \cite{Schwartz:2013pla}. The combination of these bounds leads to the inequality $m_X \lesssim 10 \text{ keV}$ for scattering to be efficient, so we should consider light dark matter. Due to the enormous neutrino energies, in the cosmic comoving frame the reaction products will be beamed to angles $\theta \lesssim \sqrt{2 m_X/E_\nu}$. Our requirement of appreciable time delay then imposes the lower bound $m_X \gtrsim 1 \text{ eV}$, though we will still need to ensure in our model that scattering is not dominated by low momentum transfer events.\footnote{In applying the scattering angle bound we are conflating the lab frame and the cosmic comoving frame, but the boost between these has negligible effect at the precision of this application.} We are thus in the regime where interactions take place at center-of-mass energy $s \simeq 2 m_X E_\nu$ and the neutrino masses do not impact the kinematics. We note also that outside of our focus on the single-scattering regime, it would be interesting to explore the multiple scattering regime which would produce similar time-delay effects while relaxing the lower bound on the dark matter mass. \footnote{Interactions of UHE neutrinos with DM in this regime were also constrained from time delay in \cite{Davis:2015rza}. The effect of interest there, however, was interactions which were so strong that neutrinos produced by the first sources would not yet have random-walked to us, and so their constraints are much weaker than the above.} 

Note that scattering also transfers energy from the neutrinos to the dark matter. In particular, the fractional energy loss in the cosmic comoving frame is $\half \left(\cos \tilde{\theta} - 1\right)$, where $\tilde{\theta}$ is the scattering angle in the center of momentum frame. Thus the larger the average number of neutrino scatterings, the more the assumed source flux must be shifted to higher energies to fit the observed diffuse background. Effects of neutrino -- dark matter scattering on the shape of the spectrum of high-energy neutrinos observed by IceCube have been discussed previously in \cite{Barranco:2010xt,Cherry:2014xra,Davis:2015rza,Reynoso:2016hjr,Arguelles:2017atb,Pandey:2018wvh,Karmakar:2018fno,Kelly:2018tyg,Huang:2018cwo}. 

We emphasize that a detailed constraint from this effect will depend, of course, on the spectrum of UHE neutrinos produced --- or at least some prior on it. We content ourselves here with an order-of-magnitude understanding, and in numerics below will show a band of $d/10 < \lambda < 10 d$, below which the effect of scattering is surely large and above which it is surely small. 

\section{A Simple Effective Model} \label{sec:model}

We will model the dark matter as being composed of a light vector boson $X^\mu$ with a Stueckelberg mass which is technically natural for any value. The fact that such vectors constitute a CDM candidate was pointed out in \cite{Nelson:2011sf}, where it was noted that inflationary misalignment can produce a condensate of light vectors which behaves like CDM at temperatures $T \lesssim T_{DM}$ where $T_{DM} = \sqrt{m_X M_{pl}}$ (further details may be found in e.g. \cite{Arias:2012az,Ringwald:2012hr,Graham:2015rva}). We will assume that there is no tree-level kinetic mixing between the dark matter and the SM photon so that the dark sector is only coupled to the SM through an effective interaction with neutrinos. 

\begin{figure}[h!]
	\centering
	\begin{subfigure}{.5\textwidth}
		\begin{center} 
			$\begin{gathered}
			\begin{fmffile}{scattering}
			\begin{fmfgraph*}(200,120)
			\fmfleft{i2,i1}
			\fmfright{o2,o1}
			\fmfset{arrow_len}{3mm}
			\fmf{fermion,label=$\nu$,label.side=left,tension=5}{i1,v1}
			\fmf{fermion,label=$\nu$,label.side=right,tension=5}{o1,v1}
			\fmf{photon,label=$X^\mu$,label.side=right,tension=5}{i2,v1}
			\fmf{photon,label=$X^\nu$,label.side=right,tension=5}{v1,o2}
			\fmfv{decoration.shape=circle,decoration.filled=shaded,decoration.size=25,label.dist=15,label.angle=80,label=$\frac{1}{M^3}$}{v1}
			\end{fmfgraph*}
			\end{fmffile}
			\end{gathered}$
		\end{center}
		\label{fig:sub1}
	\end{subfigure}%
	\begin{subfigure}{.5\textwidth}
		\centering
		\includegraphics[width=\linewidth]{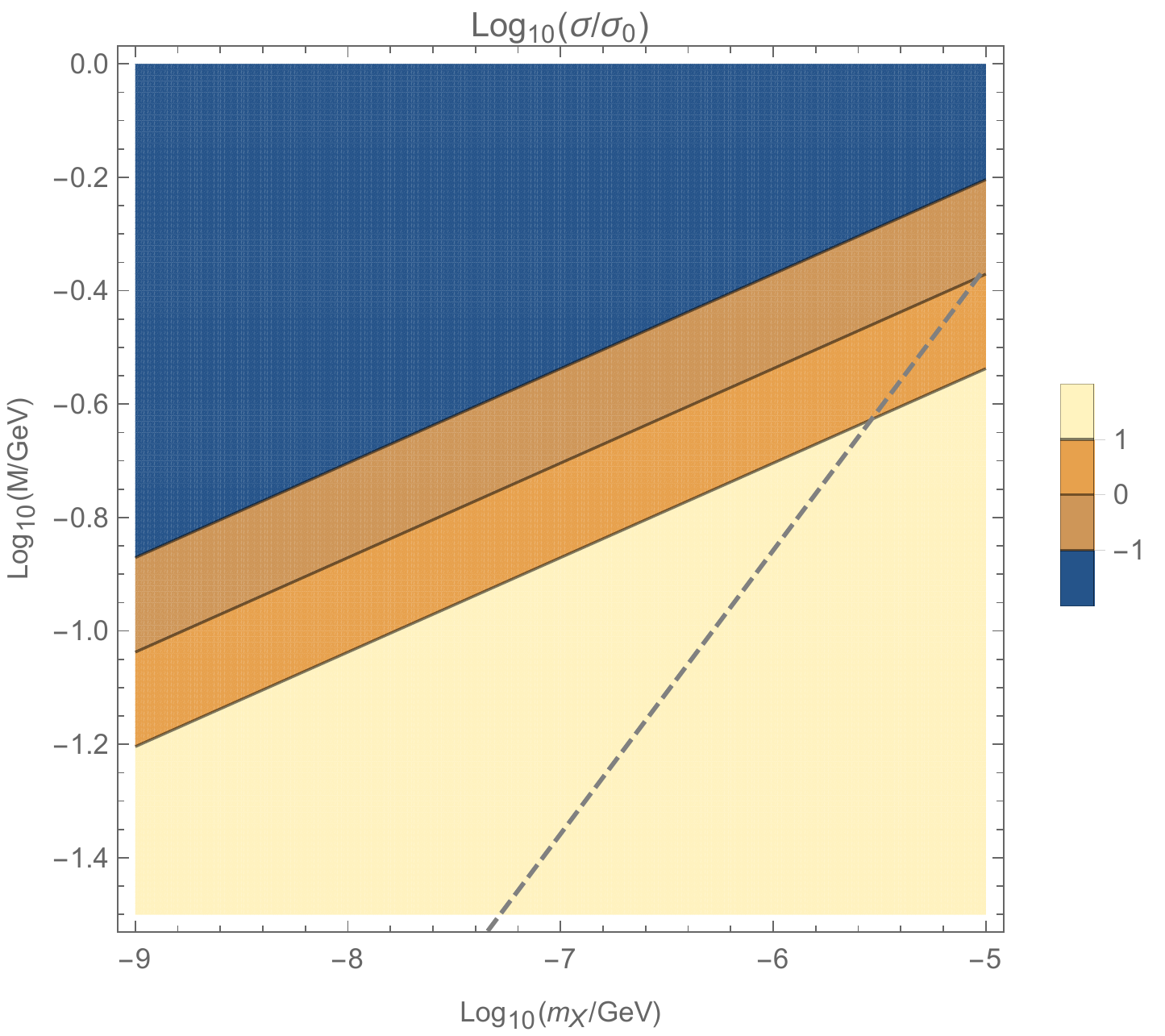}
		\label{fig:sub2}
	\end{subfigure}
	\caption{Left: Scattering of neutrinos off dark photon dark matter via the effective operator of Equation. Drawn using the \LaTeX \ package \texttt{feynMF} \cite{Ohl:1995kr}. Right: The ratio $\log_{10}(\sigma/\sigma_0)$ of the total scattering cross-section for neutrino energy $E_\nu = 300 \text{ TeV}$ to the reference cross section $\sigma_0(m_X) = 10^6 \text{ GeV}^{-3} m_X$ as a function of the dark matter mass $m_X$ and the suppression scale $M$. In the blue area scattering does not appreciably affect neutrino propagation, while in the beige area neutrinos are frequently scattered and therefore lose information about their origins. We view the orange crossover region as a conservative band of uncertainty in the exact transition location, stemming mainly from uncertainties in the source production. Below the dashed line the effective description breaks down.}
	\label{fig:crosssecs}
\end{figure}

Since we have light dark matter we are in the regime $\sqrt{s} \ll v$, with $v$ the Higgs vacuum expectation value, and so we are justified in writing an effective operator in the broken phase of electroweak symmetry. We consider the lowest-dimensional contact operator invariant under the dark Abelian symmetry,
\begin{equation}
\mathcal{L} \supset \frac{1}{M^3} \sum_k \left(\nu_k \nu_k + \nu_k^\dagger \nu_k^\dagger\right) X^{\mu\nu} X_{\mu\nu},
\end{equation}
where $\nu_k$ are SM neutrinos written in two-component notation \cite{Dreiner:2008tw}, $X^{\mu\nu}$ is the dark photon field strength, and $M$ is a suppression scale around which this effective model requires UV completion. Scattering of UHE neutrinos off dark matter proceeds through the diagram on the left of Figure \ref{fig:crosssecs}, and the cross-section is easily calculated to find
\begin{equation}
\sigma = \frac{s^2}{192 \pi M^6},
\end{equation}
neglecting very small corrections from the masses of the external states, as we are in the regime $m_\nu, m_X \ll \sqrt{s}$. On the right of Figure \ref{fig:crosssecs} we show the ratio of the cross section predicted by this model to that required for significant scattering, as discussed in Section \ref{sec:requirements}. We focus on the region of parameter space in which scattering is marginally efficient, which is that relevant for the single-scattering regime, and which we will also find in Section \ref{sec:constraints} is favored by cosmology. The region of parameter space in which scattering is efficient may either be ruled out by the correlation of a UHE neutrino with an electromagnetic signal, or may be responsible for the dearth of such observations. The effective description violates the unitarity bound below the dashed line and so must break down by then at the latest. Scattering can continue to be efficient in the UV description, but the precise dependence of the cross-section on energy will depend on the new degrees of freedom which come in around the scale $M$.

We note that the differential scattering rate through this effective operator peaks at $t \sim -s$, so scattering is most efficient near $\cos \tilde{\theta} \sim -1$ when this description is valid, with $\tilde{\theta}$ the scattering angle in the center-of-mass frame. Thus the source spectrum implied by the observed diffuse UHE neutrinos increasingly blueshifts for $d/\lambda \gg 1$. So at some point such models should be considered ruled out instead by the existence of the diffuse neutrino flux and upper limits on the plausible source energies. However, given the large uncertainties on the source spectra, there is still a wide swath of parameter space which remains probed only by neutrinos' correlations with sources. If this effective process is UV completed above scales around $M$ into $t-$channel scattering, momentum transfer will then be peaked at $t \sim -M^2$ for $s > M^2$, and scattering will no longer be peaked at large momentum transfer.

The question of whether the typical time delay is large enough to decorrelate neutrinos from their sources is the converse question to that of energy loss --- it is low momentum transfer events which are forward scattered and do not pick up a large time delay. Since scattering is dominated by large momentum transfer interactions here, it is clear that this IR scattering cutoff has no effect and all scatterings cause high delays. This analysis neglects the fact that not all DM is homogeneously distributed and that we have a nearby region of above-average dark matter density --- namely, the galactic center. However, due to its proximity to us, scattering events in the galactic center do not cause appreciable time delay. This simply means we are correct not to consider this inhomogeneity in estimating the scattering rate.

This inhomogeneity does cause another effect by which IceCube data may be used to probe this scenario, in the case that no associations of UHE neutrinos with sources are observed. In the presence of neutrino -- dark matter scattering the average energy of detected neutrinos will become anisotropic, as neutrinos we detect from the direction of the galactic center will have scattered a greater number of times. Note, importantly, that this effect is independent of the assumed source spectrum, and so is a smoking gun signature of neutrino -- dark matter scattering. As this effect requires large statistics to overcome the energy- and directional-uncertainties of the IceCube detections, we leave this analysis for future work.\footnote{This suggestion resembles the analysis of \cite{Arguelles:2017atb}, but differs in assumptions about the source spectrum. Their conclusion of an anisotropic \textit{rate} of neutrino detections (in particular a dearth of detections toward the galactic center) is dependent upon their assumption of a power-law source spectrum --- inspired by fits to the observed spectrum --- which results in many scattered neutrinos falling out of the observable energy window of IceCube. However, the GRB spectra of UHE neutrinos are believed to be `bump-shaped' and there is considerable phenomenological freedom to translate the peak of the spectrum  (see e.g. \cite{He_2012,Zhang:2012qy,Meszaros:2015krr}), so from our perspective the observed spectrum may be the result of significant processing by dark matter interactions. It would be interesting to repeat the analysis of \cite{Arguelles:2017atb} in a way that accounted for these considerable uncertainties in the source spectrum and so constrained solely the energy anisotropy. 
Note also that \cite{Davis:2015rza} find that neutrino -- dark matter scattering \textit{enhances} the rate of neutrinos detected from the galactic center, where they have used source spectra inspired by models for hypernova remnants. This disagreement underscores the need to disambiguate the effects from the assumed source spectrum and the effects from the neutrino -- dark matter scattering.
}

\section{Constraints} \label{sec:constraints}

An unavoidable effect of neutrino -- dark matter scattering is high energy neutrino annihilation with the cosmic neutrino background into dark photons, which is given by crossing the scattering diagram. However, these annihilation events occur at significantly lower center-of-mass energies than does scattering, since $m_\nu \ll m_X$ \cite{Aghanim:2018eyx}. Since $\sigma \propto s^2$ this suppression is large, and the requirement of inefficient annihilations merely tells us we must not be in the regime where scattering occurs too often ($d \ll \lambda$), which we knew already from the implied blueshifting of the source spectra, as remarked above.

The sole other observation of neutrinos coincident with an astrophysical source is the supernova SN1987A. However, those are neutrinos from the core-collapse itself, before the point at which a GRB could be produced by a fireball and beamed toward us at far higher energies \cite{Woosley:2006fn}. Thus these neutrinos all have energies ${\sim} 10 \text{ MeV}$ \cite{Hirata:1987hu}, and so the center-of-mass energy with which they interact with dark matter is a factor of ${\sim} 10^4$ smaller. Furthermore, this supernova took place at a distance of only $d \sim 50 \text{ kpc}$, though our line-of-sight with its source likely contains a larger average line density of dark matter than does the intergalactic medium. However, scattering at these energies may easily be seen to be negligible from Figure \ref{fig:crosssecs}. Since the dark matter mass and the neutrino energy only enter through the combination $s=2m_X E_{\nu}$, the effect of the lower energy is to shift the plot 8 decades to the right. As a result, in the region of parameter space where interactions are efficient for UHE neutrinos they are inefficient for supernova neutrinos, as long as we are not in the regime where scattering occurs very often.

Interactions between neutrinos and dark matter may also have interesting cosmological effects. These have been used to constrain interactions of thermal DM with neutrinos \cite{Boehm:2000gq,Mangano:2006mp,Nollett:2014lwa,Schewtschenko:2014fca,Escudero:2015yka,Bringmann:2016ilk,Campo:2017nwh} (or to suggest that such interactions may solve cosmological problems \cite{Aarssen:2012fx,Bertoni:2014mva,Cherry:2014xra,Bringmann:2016ilk,DiValentino:2017oaw}), but we are unaware of any studies on the cosmological effects of light, non-thermal CDM interactions with neutrinos. A full discussion and calculation of constraints would require numerical solution of the Boltzmann equations and is beyond the scope of this work, but we may perform a few checks on the likely effects. 

A consistency requirement is that the dark photon dark matter is not brought into thermal equilibrium in the early universe. Appreciable astrophysical effects require $\lambda_\nu(E_\nu) < d$ today, with $d \sim 10^{41} \text{ GeV}^{-1}$, and we may frame the condition of not thermalizing as roughly that the mean free path of dark matter in the early universe is not below the Hubble length $\lambda_{DM}(T) >H^{-1}(T)$, with $T$ the temperature. We may relate the two as $\lambda_{DM}(T) \sim \frac{n_{DM}(T_0)}{n_\nu(T)}\left(\frac{E_\nu}{T}\right)^2 \lambda_\nu(E_\nu)$, using the fact that our cross section scales as $\sigma \propto s^2$. Since $\lambda_{DM}$ depends more strongly on temperature than does $H$, we must check that the DM is not brought into kinetic equilibrium at the highest accessible temperatures. In the regime of single scattering, the large interaction cross-section required would lead to thermalization at high temperatures. We thus here restrict to scenarios of low-scale reheating, though it would be interesting to explore the multiple-scattering regime in which lower dark matter masses would be allowed. Taking the reheating temperature all the way down to $T_R \gtrsim 5 \text{ MeV}$, we find $\lambda_\nu(E_\nu) \gtrsim 10^{37} \text{ GeV}^{-1} \left(\frac{m_X}{10^{-9} \text{ GeV}}\right)$. Neutrino -- dark matter interactions must thus not be \textit{too} much stronger than would cause decorrelation of UHE neutrinos from their sources, as we have already found above. We emphasize that numerical solution of the Boltzmann equations is required to fully understand these constraints, and our analysis here is just to show that this parameter space is plausible.

We must also consider whether the neutrino phase space distribution will be modified by the dark matter interactions. The mean free path of neutrinos after they have decoupled from the SM bath at $T_{dec}$ may similarly be related to the mean free path today, and one finds for $T_{dec}\sim 1 \text{ MeV}$, $\lambda_\nu(T_{dec}) \sim 10^{-12} \lambda_\nu(E_\nu)$ while $H^{-1}(T_{dec}) \sim 10^{25} \text{ GeV}^{-1}$. So again as long as these interactions are not \textit{too} strong, they have no effect on neutrinos in the early universe. 

The above considerations all depend only on the external scattering states. The overall picture is that, while interactions between leptons and dark matter are probed by a variety of different searches and observations, these take place in very different kinematic regimes from astrophysical ultra-high-energy neutrino scattering. It is only in the early universe, when large number densities may overcome the drop-off in cross-section, that constraints on this regime of neutrino -- dark matter interaction may be found. Interactions which proceed through higher-dimensional operators would be less-constrained from cosmological observations than this model. 

\section{Conclusion} \label{sec:conclusion}

In this work we have taken a bipartite approach to the study of ultra-high-energy neutrino interactions with dark matter. On the one side we have pointed out that the presence of a large scattering rate may solve the persistent astrophysical mystery of why no neutrinos have been observed in association with a gamma ray burst. And we have described that searches for anisotropic energies of UHE neutrinos would provide a smoking gun signature which may be used to confirm this hypothesis even in the presence of large astrophysical source uncertainties. On the other side we have pointed out that the association of a neutrino with the blazar TXS 0506+056 supports the conclusion that neutrinos free-stream at high energies, which is qualitatively new information. By continuing to search for neutrinos coincident with high-energy electromagnetic astrophysical events we will be able to distinguish these two possible conclusions. 

We have here given one simple effective model in which elastic scattering of neutrinos off dark matter may be large, but our qualitative conclusion about the effects of neutrino -- dark matter scattering are more general. It would be interesting to explore UV completions of this effective operator or others, including with a variety of dark matter candidates. The large cross-section required may present a challenge, as suppression by the neutrino masses must be avoided, but this avenue of model-building will become more appealing if further associations of UHE neutrinos with EM sources are not found in the coming years. It would also be useful to embed this idea in a more-complete picture of neutrino physics beyond the Standard Model - perhaps to address the variety of low-energy evidences for light sterile neutrinos (for reviews see \cite{Abazajian:2012ys,Kopp:2013vaa,Palazzo:2013me,Drewes:2013gca,Gariazzo:2015rra}) or to further study the extent to which neutrino -- dark matter interactions may help resolve various tensions in $\Lambda \text{CDM}$ (e.g. \cite{Cherry:2014xra,Aarssen:2012fx,Bertoni:2014mva,DiValentino:2017oaw}).
Most ambitiously, a model in which the dark matter abundance is a result of its interactions with neutrinos would be especially compelling.

We have also drawn attention to the deficit of work on cosmological interactions of neutrinos with non-thermal dark matter, further study of which would help delimit the allowed parameter space for UHE neutrino interactions. Finally, we have limited our analysis of the effects of scattering off dark matter to considering a single UHE neutrino energy and a single scattering. There is clearly more to be gained by performing a more complete analysis, which would entail a full examination of the theorized energy distributions of UHE neutrinos from sources, how this distribution could be modified by scattering off galactic and intergalactic dark matter, and how well such an effect could match the observed diffuse UHE neutrino flux.

\acknowledgments

It is a pleasure to thank Andr\'{e} de Gouv\^{e}a, John Mason, Dave Sutherland, and especially Nathaniel Craig for helpful discussions and for comments on a draft of this work. We are grateful to the authors of \cite{Arguelles:2017atb} for a dialogue about their paper, and to an anonymous referee for drawing our attention to an error in a previous version of this manuscript. This work is supported in part by the US Department of Energy under the grant DE-SC0014129. 

\bibliography{nudm}
\bibliographystyle{JHEP}

\end{document}